\pdfoutput=1

\documentclass[11pt,twoside,a4paper,cmspaper,final,collab]{cms-tdr}
\def\svnVersion{245654}\def\cmsCernNo{2013-167}\def\cmsCernDate{\today}\def\cmsMessage{Published in Physics Letters B as \href{http://dx.doi.org/10.1016/j.physletb.2014.05.055}{\doi{10.1016/j.physletb.2014.05.055.}}}

\begin{document}\cmsNoteHeader{BPH-11-026}

\hyphenation{had-ron-i-za-tion}
\hyphenation{cal-or-i-me-ter}
\hyphenation{de-vices}

\RCS$Revision: 239111 $
\RCS$HeadURL: svn+ssh://svn.cern.ch/reps/tdr2/papers/BPH-11-026/trunk/BPH-11-026.tex $
\RCS$Id: BPH-11-026.tex 239111 2014-04-29 22:16:15Z alverson $
\newcommand{\massdiffoneoned}{\ensuremath{1051.4 \pm 2.7\stat\MeV}\xspace}
\newcommand{\massdifftwooned}{\ensuremath{1220.2 \pm 4.1\stat\MeV}\xspace}
\newcommand{\widthoneoned}{\ensuremath{30 \pm 12\stat\MeV}\xspace}
\newcommand{\widthtwooned}{\ensuremath{35 \pm 17\stat\MeV}\xspace}
\newcommand{\massdiffone}{\ensuremath{1051.3 \pm 2.4\stat\MeV}\xspace}
\newcommand{\massdifftwo}{\ensuremath{1217.1 \pm 5.3\stat\MeV}\xspace}
\newcommand{\massonewithsys}{\ensuremath{4148.0 \pm 2.4\stat \pm 6.3\syst\MeV}\xspace}
\newcommand{\masstwowithsys}{\ensuremath{4313.8 \pm 5.3\stat \pm 7.3\syst\MeV}\xspace}
\newcommand{\widthone}{\ensuremath{28 _{-11} ^{+15}\stat\MeV}\xspace}
\newcommand{\widthonewithsys}{\ensuremath{28 _{-11} ^{+15}\stat \pm  19\syst\MeV}\xspace}
\newcommand{\widthtwo}{\ensuremath{38 _{-15} ^{+30}\stat\MeV}\xspace}
\newcommand{\widthtwowithsys}{\ensuremath{38 _{-15} ^{+30}\stat \pm  16\syst\MeV}\xspace}
\ifthenelse{\boolean{cms@external}}{\providecommand{\suppMaterial}{the supplemental material [URL will be inserted by publisher]\xspace}} {\providecommand{\suppMaterial}{Appendix~\ref{app:suppMat}\xspace}}
\ifthenelse{\boolean{cms@external}}{\providecommand{\suppMaterialSecond}{the supplemental material\xspace}} {\providecommand{\suppMaterialSecond}{Appendix~\ref{app:suppMat}\xspace}}

\cmsNoteHeader{BPH-11-026} 
\title{
Observation of a peaking structure in  the $\JPsi\phi$ mass spectrum from $\PBpm \to \JPsi \phi \PKpm$ decays}

\date{\today}

\abstract{

A peaking structure in the $\JPsi\phi$  mass spectrum near
threshold is observed in $\PBpm \to \JPsi \phi \PKpm$
decays, produced in pp collisions at $\sqrt{s} = 7$\TeV collected with the CMS detector at the LHC.  The data sample,
selected on the basis of the dimuon decay mode of the
$\JPsi$, corresponds to an integrated
luminosity of 5.2\fbinv.  Fitting the structure to an $S$-wave relativistic
Breit--Wigner lineshape above a three-body phase-space nonresonant component
gives a signal statistical significance exceeding five standard deviations.
The fitted mass and width values are
$m=\massonewithsys$ and $\Gamma=\widthonewithsys$, respectively.
Evidence for an additional peaking structure at higher $\JPsi \phi$ mass
is also reported.

}

\hypersetup{%
pdfauthor={CMS Collaboration},%
pdftitle={Observation of a peaking structure in  the J/psi phi mass spectrum from   B+/- to J/psi phi K+/- decays},
pdfsubject={CMS},%
pdfkeywords={CMS, physics}}

\maketitle 

\section{Introduction}

The discovery of  new charmonium-like
states~\cite{Choi:2003ue,Acosta:2003zx,Abazov:2004kp,Aubert:2004ns,Abe:2004zs,Aubert:2007vj} over the last
decade poses a  challenge to the conventional quark model.
Many explanations,
such as charmed hybrids, tetraquarks, and molecular states, have been proposed for these new entities, but
their nature remains a puzzle~\cite{Olsen:2010ex,Brambilla:2010cs}.
In 2009, the CDF Collaboration  reported evidence for a narrow structure, which they
called $\mathrm{Y}(4140)$, near the $\JPsi\phi$ threshold in
 $\PBpm\to \JPsi\phi \PKpm$ decays~\cite{Aaltonen:2009tz}.
This structure, if confirmed as a new resonance,  would be a candidate for an exotic
meson~\cite{Liu:2009ei,Mahajan:2009pj,Branz:2009yt,Albuquerque:2009ak,Liu:2009iw,Ding:2009vd,Zhang:2009st,Wang:2009ue,Molina:2009ct}.
The Belle Collaboration searched for the $\mathrm{Y}(4140)$ through the same
 $\PBpm$ decay channel~\cite{Uglov:2010wc} and in the two-photon process
$\gamma\gamma\to \JPsi\phi$~\cite{Shen:2009vs}, but did not confirm it.
Using the same $\PBpm$ decay channel,
the LHCb Collaboration recently reported finding no evidence for such a state,
in disagreement with the CDF result~\cite{Aaij:2012pz}.

In this Letter, a  study of the $\JPsi\phi$ mass spectrum from
 $\PBp\to \JPsi\phi \PKp$
decays is reported, where charge conjugate decay modes are implied throughout.
The data were collected in 2011 with the Compact Muon Solenoid (CMS) detector from proton-proton collisions
at the Large Hadron Collider (LHC) operating at a center-of-mass energy of 7\TeV
and   corresponding to an integrated luminosity of
$5.2 \pm 0.1$\fbinv~\cite{lumi2011}.

A detailed description of CMS can be found
elsewhere~\cite{cmsdet}.
The central feature of the CMS
apparatus is a superconducting solenoid, 13\unit{m} long with a 6\unit{m} internal diameter,
which provides an axial magnetic field of 3.8\unit{T}.  Within the field
volume is the silicon tracker, which consists
of a pixel-based detector in the inner region and layers
 of microstrip detectors in the outer region.
Charged-particle trajectories are
measured with the silicon tracker, covering $0  < \phi  \leq  2\pi$ in
azimuth and $\abs{\eta} <  2.5$, where the pseudorapidity $\eta$ is defined as
$-\ln(\tan[\theta/2])$ and $\theta$ is the polar angle of the trajectory
of the particle with respect to the counterclockwise-beam direction.
Muons are detected in the pseudorapidity range $\abs{\eta}<2.4$ by three types
of gas-ionization detectors embedded in the steel flux-return yoke of the magnet: drift tubes in the barrel,
cathode strip chambers in the endcaps, and resistive-plate chambers in both the barrel
and endcaps. The strong magnetic field and excellent position resolution
of the silicon tracker enable the transverse momentum ($\pt$) of a muon matched to a reconstructed
track  to be measured with a resolution of approximately  0.7\% for $\pt$ of 1\GeV.
The pixel detector, with its excellent spatial resolution
and low occupancy,  enables the separation of $\PBp$-decay vertices from the
primary interaction vertex.

Monte Carlo (MC) simulated data were created using
\PYTHIA6~\cite{Pythia} for the particle production,
\EVTGEN~\cite{EvtGen} for the particle decays, and
\GEANTfour~\cite{Geant4} for tracing the particles through a detailed
model of the detector.  These samples were created with the appropriate
conditions for the data analyzed, including the effects of alignment,
efficiency, and number of simultaneous pp collisions.

\section{Event selection}

Events are chosen   using a two-level trigger system.
The first level, composed of custom hardware processors,
uses information from the muon detectors to select  dimuon candidates.
The high-level trigger (HLT) runs a special version of the offline
software  code on a processor farm to
select events with nonprompt $\JPsi$ candidates coming from
the decays of $\PB$ mesons.

Events containing $\JPsi$ candidates are
selected by the HLT dimuon trigger.
Because of the increasing LHC instantaneous luminosity, there are
two configurations of the HLT, corresponding to two running
periods and two distinct data sets.
For both data sets, the following requirements are already applied  with the HLT\@.
The dimuon  $\pt$  is required to be
greater than 6.9\GeV, the two muons must be oppositely charged
and form a  three-dimensional (3D) vertex with a $\chi^2$ probability
greater than 0.5--10\%, depending on the running period.
The resulting $\JPsi$ vertex must be displaced
from the average interaction point (beamspot) in the transverse plane by at least three times
its uncertainty, which is the sum in quadrature of the secondary-vertex
uncertainty and the beamspot size in the transverse plane.
The cosine of the angle between the transverse projections of
the line joining the beamspot and dimuon vertex and the dimuon momentum
direction must exceed 0.9.
For the later data set, there is
an additional requirement that the $\pt$ of each muon be greater than 4\GeV.
In the final selection of \cPJgy~candidates,
the  dimuon $\pt$  is required to be
greater than 7\GeV,  the $\chi^2$ probability of the
dimuon vertex is demanded to be greater than 10\%,
and the reconstructed dimuon invariant mass
must be within 150\MeV of the $\JPsi$ mass~\cite{Beringer:1900zz}.

The $\PBp\to \JPsi \phi \PKp$ candidates are reconstructed by combining three additional
charged-particle tracks that are consistent with originating from the
displaced $\JPsi$ vertex and have
a total charge of ${\pm}1$.
These tracks are assigned the kaon mass and this mass is used in
accounting for the effects of energy loss and multiple-scattering.
We do not apply a mass constraint on the $\phi$ candidate  because
our experimental $\PKp\PKm$ mass resolution (1.3\MeV) is
less than the $\phi$ meson natural width (4.3\MeV).
The  $\pt$ of all kaon tracks are required to be greater than 1\GeV.
Only tracks that pass the
standard CMS quality requirements~\cite{Khachatryan:2010pw} are used.
The five tracks, with the $\Pgmp\Pgmm$ invariant mass constrained to the   $\JPsi$ mass,
are required to form a good 3D vertex with a $\chi^2$
probability greater than 1\%.
There are two $\PKp\PKm$ combinations from the three charged kaon
tracks, and we use
the lower invariant mass as the $\phi$ candidate;
MC simulations of the $\PBp$ decay predict that the $\phi$ signal
from the other combination is negligible, which is verified in the data.
The reconstructed  $\PKp\PKm$ invariant mass  must satisfy
$1.008\GeV<m(\PKp\PKm)<1.035$\GeV to be considered as a  $\phi$ candidate.
These selection requirements were designed to maintain high efficiency
for $\PBp$ decays and were fixed before the $\JPsi\phi$ mass spectrum in
data was examined.

\section{Results}

\begin{figure*}[htb]
  \begin{center}
   \includegraphics[width=\linewidth]{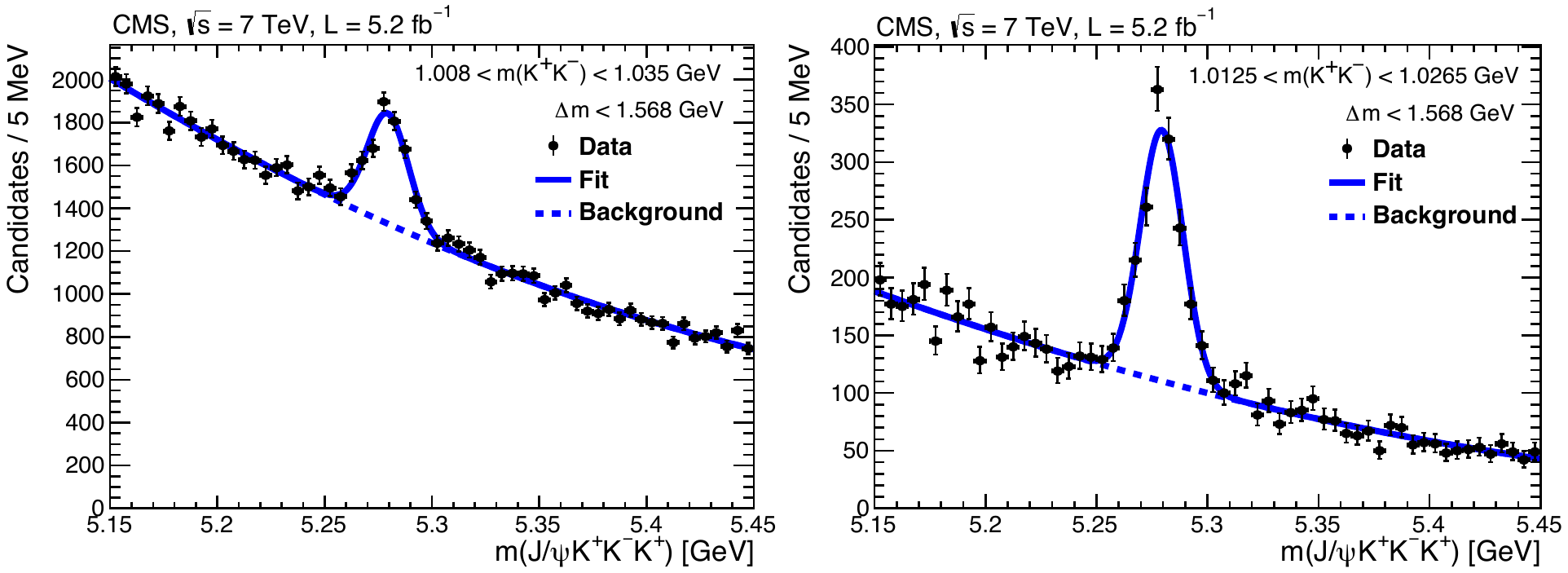}
    \caption{
The $\JPsi \phi \PKp$ mass distribution with the standard event selection
(left) and the tighter requirements (right).  The solid curves show the
result of fitting these distributions to a Gaussian signal and a
second-degree polynomial background while the dashed curves show the
background contribution.
    }
    \label{fig:bmass}
  \end{center}
\end{figure*}

The invariant-mass spectrum of the selected $\JPsi\phi \PKp$ candidates  is
shown in the left plot of Fig.~\ref{fig:bmass} for a mass difference
$\Delta m \equiv m(\Pgmp\Pgmm\PKp\PKm)-m(\Pgmp\Pgmm) <1.568$\GeV.
We only investigate candidates with
$\Delta m< 1.568\GeV$ because of possible  background from  $\PBzs\to \Pgy \phi \to \JPsi\pi^+\pi^-\phi$
at higher values, as discussed below.
The invariant-mass spectrum is fit with a Gaussian signal function and
a second-degree  polynomial background function.
The fit returns a $\PBp$ mass of $5.2796\pm0.0006\stat$\GeV, which agrees with
the nominal value~\cite{Beringer:1900zz}, and a Gaussian
width of $9.6 \pm 0.7\stat$\MeV, which is consistent with the prediction
from the MC  simulation.
The $\PBp$ yield is  $2480\pm160\stat$ events,
which  is   the world's largest  $\PBp\to \JPsi\, \phi \PKp$
sample. The combined $\PBp$ yield is  $2340\pm120\stat$ events when each data set is fit
with two  Gaussian signal functions and the width of each function is fixed to the prediction
from MC simulation.
Approximately 5\% of the selected events have
more than one $\PBp$ candidate within 1.5 times our mass resolution $(\sigma)$ of
the  $\PBp$ mass; all candidates are kept.

The right plot in Fig.~\ref{fig:bmass} displays
the $\JPsi\, \PKp\,\PKm\,\PKp$
invariant-mass distribution after making
the following tighter requirements:
the $\pt$ of the kaons must be greater than   1.5\GeV, the $\PBp$ vertex probability must be greater than 10\%,
the $\PBp$ vertex  must be displaced  from the primary vertex  in the transverse plane
by at least seven times its uncertainty, and  $m(\PKp\PKm)$ must be within 7\MeV of the
$\phi$ meson mass~\cite{Beringer:1900zz}.
With these requirements, 40\% of the $\PBp$ candidates are retained,  while the background is reduced by
more than a factor of ten. This sample of cleaner signal candidates is used as  a cross-check of the results
obtained by employing the  background-corrected $\JPsi\phi$ mass spectrum,
as described below.
With the exception of this cross-check, all results are obtained with the
less-restrictive  criteria.

\begin{figure}[!htb]
  \begin{center}
   \includegraphics[width=\linewidth]{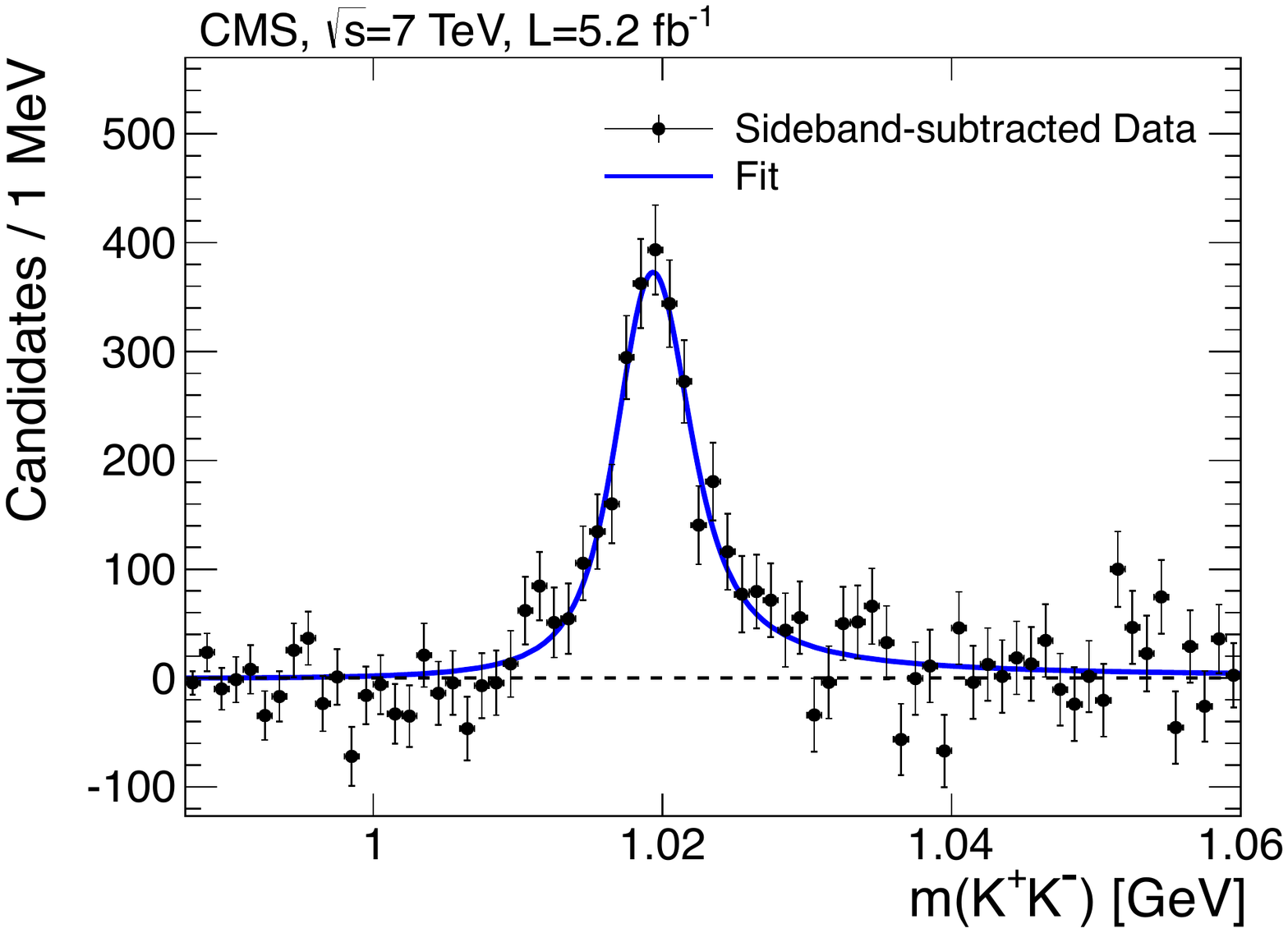}
    \caption{
The $\PBp$ sideband-subtracted  $\PKp \PKm$
invariant-mass distribution for  $\JPsi \PKp\PKm  \PKp$
candidates  within $\pm 3 \sigma$ of the nominal $\PBp$  mass.
The solid curve is the result of the fit described in the text.
The dashed line shows the zero-candidate baseline.
    }
    \label{fig:phimass}
  \end{center}
\end{figure}

Figure~\ref{fig:phimass} shows the $\PKp \PKm$  invariant-mass
distribution for $\JPsi\, \PKp\,\PKm\,\PKp$ candidates
that have an invariant mass within $\pm 3\sigma$ of the $\PBp$ mass.
We define  events in the range  $[-12,\,-6]\sigma$ and $[6,\,12]\sigma$ of
the $\PBp$ mass as sidebands.
The $\phi$ mass restriction has been removed and
a sideband subtraction has been performed in Fig.~\ref{fig:phimass}.
We fit this distribution
to a $P$-wave relativistic
Breit--Wigner (BW) function
convolved with
a Gaussian resolution function. The width of the Gaussian is fixed to 1.3\MeV,
obtained from MC simulation. The fit has a  $\chi^2$ probability of 23\%  and returns
a mass of $1019.4\pm 0.1$\MeV and a width
of $4.7\pm   0.4$\MeV,
consistent with  the $\phi$ meson~\cite{Beringer:1900zz}.
The good fit to only a $\phi$ component in Fig.~\ref{fig:phimass}
indicates that after the $\JPsi$ and $\phi$ mass requirements are made
and the combinatorial background is subtracted,
the $\PBp\to \Pgmp\Pgmm \PKp \PKm \PKp$
candidates are consistent with being solely
$\JPsi\phi \PKp$, with negligible
contribution from $\JPsi \mathrm{f}_0(980) \PKp$ or
nonresonant $\JPsi \PKp \PKm\PKp$.

As seen in Fig.~\ref{fig:bmass},
there are two main components to  the $\JPsi\phi \PKp$  invariant-mass
spectrum: the $\PBp$ signal and a smooth background.
Possible contributions from other $\PB$-hadron decays are examined using MC
simulations of  inclusive
$\PBp$, $\PBz$, and \PBzs decays. Based on this study,
the mass-difference  region ($\Delta m>1.568$\GeV) is
excluded from the analysis
to avoid potential background
from $\PBzs \to \Pgy \phi \to \JPsi\pi^+\pi^-\phi$ decays,
where one pion is assumed to be a kaon and the other  is not reconstructed.

\begin{figure}[!htb]
  \begin{center}
   \includegraphics[width=\linewidth]{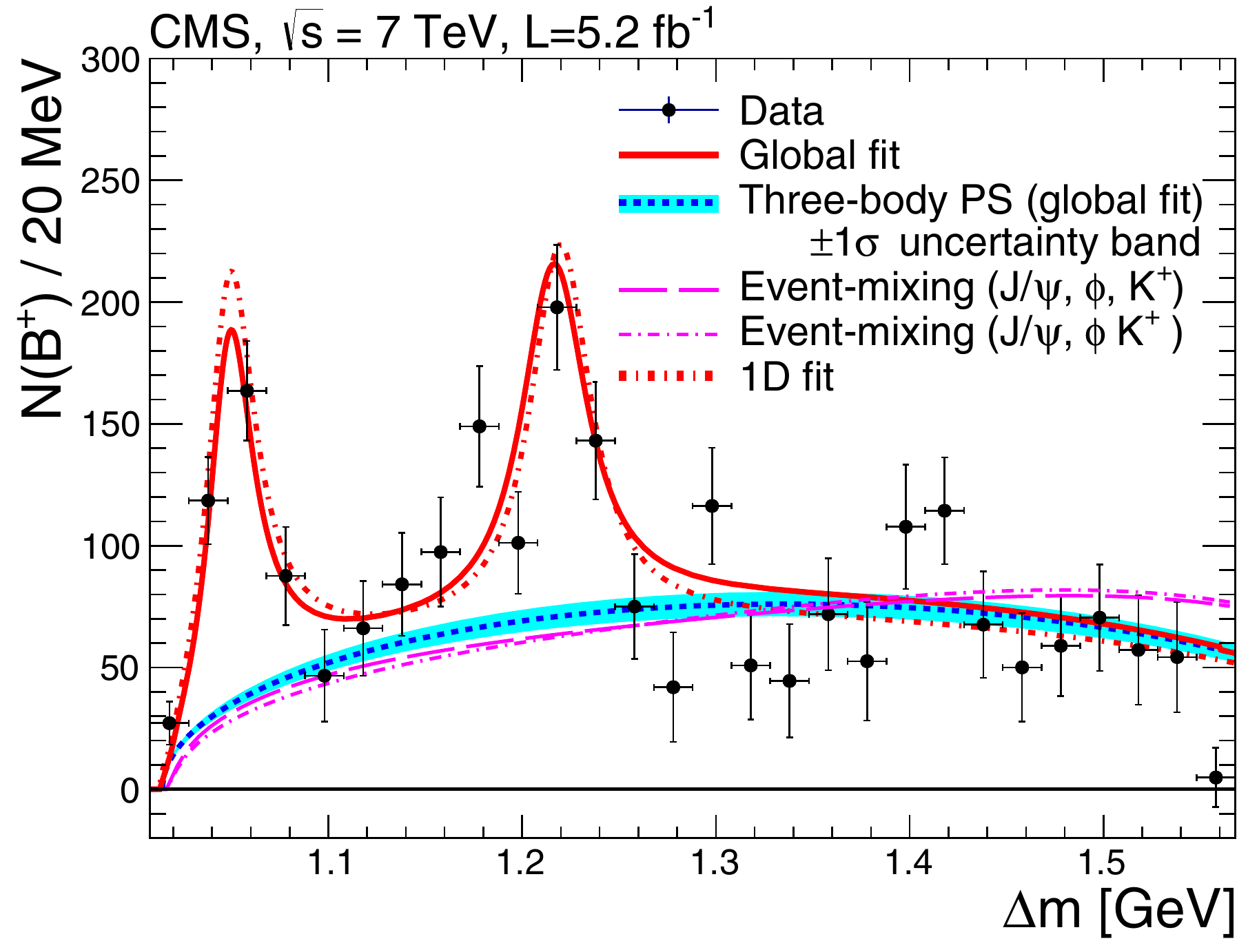}
    \caption{
The number of  $\PBp\to \JPsi \phi \PKp$  candidates as a function
of $\Delta m = m(\Pgmp\Pgmm\PKp\PKm)-m(\Pgmp\Pgmm)$.
The  solid curve is the  global unbinned maximum-likelihood fit  of the data, and the
dotted curve is the  background contribution assuming three-body PS.
 The  band is the $\pm 1\sigma$
uncertainty range for the background  obtained from the global fit.
The dashed and dash-dotted curves are background curves obtained from two different event-mixing
procedures, as described in the text, and normalized to the number of three-body PS
background events.
The short dashed curve is the 1D fit to the data.
    }
    \label{newfig3}
  \end{center}
\end{figure}

To investigate the $\JPsi\phi$ invariant-mass distribution,
rather than fitting the distribution itself with its large combinatorial background,
the  $\JPsi\phi \PKp$ candidates are divided into 20\MeV-wide
$\Delta m$ intervals,
and the  $\JPsi\phi \PKp$ mass distributions for each interval are fit to extract
the $\PBp$ signal yield in that interval.
We use a second-degree  polynomial  for the combinatorial background and two
Gaussians for the $\PBp$ signal.
The fit is performed separately
for each data set.
The mean values of the two Gaussians are fixed to the   $\PBp$ mass~\cite{Beringer:1900zz},
and the width values of the  Gaussians, as well as their relative ratio,
are fixed to the values obtained from
MC simulation  for each specific $\Delta m$ interval in each data set.
The results of all the fits are good descriptions of the data distributions
with an average $\chi^2$
per degree of freedom (dof)  close to 1. The resulting $\Delta m$ distribution
for the combined  data sets  is shown in Fig.~\ref{newfig3}.
Two peaking structures are observed above the simulated phase-space (PS) continuum
distribution shown by the dotted line.

Results obtained from both data sets  are consistent.
We have checked that events with multiple $\PBp$ candidates
do not artificially enhance the two structures.
The total number  of  $\PBp$ signal events
in the $\Delta m$ intervals below 1.568\GeV
is $2320\pm110\stat$, which is consistent with the total number of $\PBp$
candidates estimated from the mass spectrum in Fig.~\ref{fig:bmass}.

A full study of the $\JPsi\phi$ resonant pattern in the
$\PBp \to \Pgmp\Pgmm \PKp \PKm \PKp$
decay via an amplitude analysis of the five-body decay  would require a data sample at
least an order of magnitude larger than is currently available,
as well as more precise information on possible $\phi \PKp$
or $\JPsi \PKp$ resonances that may contribute to this decay.
Instead, the $\Delta m$ distribution is studied, since it is
related to the projection of the two-dimensional (2D)
$\JPsi \phi \PKp$ Dalitz
plot onto the $m^{2}(\JPsi \phi)$ axis.

Before fitting the  $\Delta m$ distribution, it  must be corrected for
the relative detection and reconstruction efficiencies of the candidate events.
Since no branching fractions are being determined, only the relative
efficiency over the Dalitz plot is required.
If a possible  $\phi \PKp$ or  $\JPsi \PKp$ resonance did exist,
the density of events would depend on the quantum numbers of the resonance and on the
interference of the two structures with the  possible resonance.
Ignoring these possible interference effects, the MC simulation is used to
determine the efficiency over the $m^2(\phi \PKp)$
vs.\ $m^2(\JPsi\phi)$ Dalitz plot,  assuming a
PS distribution for the three-body decay
$\PBp \to \JPsi \phi \PKp $.
The $\JPsi$ and $\phi$ vector meson decays  are simulated  using their known
angular distributions according to the VLL and VSS model in \EVTGEN, while we assume there is no
polarization for the two vectors. The PS MC simulation
is reweighted assuming either transverse or longitudinal $\JPsi$ and $\phi$ polarization.
The effect of either polarization  is found to be   negligible.
The measured efficiency is fairly uniform, varying by less than 25\% over
the entire allowed three-body PS.
Assuming a uniform PS distribution,
the efficiency
for each  $\Delta m$ bin  is taken to be  the average
of the efficiencies over the full
kinematically allowed  $m( \phi \PKp )$ range.
To estimate the systematic uncertainty in the efficiency caused by
its  dependence on the unknown quantum numbers of the structures, and
hence on their unknown decay angular distributions,
the efficiency is evaluated under the assumption of both a $\cos ^{2} \theta$
and  $\sin ^{2}\theta $ dependence,
where $\theta$ is the helicity
angle, defined as the angle in the $\JPsi \phi $ rest frame
between the direction of the boost from the
laboratory frame and the  $\JPsi$ direction.
Since the efficiency tends to be
lower towards the edge of the Dalitz plot,
the $\cos ^{2} \theta$ dependence gives a lower average efficiency than the default efficiency,
while the $\sin ^{2}\theta $ dependence gives a slightly higher average efficiency.
This variation (10\%) is taken as the
systematic uncertainty in the
efficiency from our lack of knowledge of the quantum numbers of
the structures and the effects of interference with
possible two-body resonances.

We investigate the possibility that the two structures in the
$\Delta m$ distribution are caused by reflections from resonances
in the other two-body systems,
$\JPsi \PKp$ and $\phi \PKp$.
Such reflections are well known in the two-body systems from other
three-body decays because of kinematic constraints.
There are candidate states that decay
to $\phi \PKp$~\cite{Beringer:1900zz},
although they are not well established.
These could potentially produce reflected structures in the $\JPsi\phi$ spectrum.
In particular, a $D$-wave contribution to $\PKm\mathrm{p}$ scattering in the mass region
around 1.7--1.8\GeV has  been reported by several fixed-target
experiments~\cite{Daum:1981hb,Armstrong:1982tw,Aston:1993qc}.
This is interpreted  as two interfering broad $J^P=2^-$
 resonances, labeled  $\mathrm{K}_2(1770)$ and $\mathrm{K}_2(1820)$, with widths in the
 range  200--300\MeV. These resonances at relatively low $\phi \PKp$
mass cannot
 affect the $\JPsi\phi$ structure near  threshold, but could contribute to
 the second $\JPsi\phi$ structure near $\Delta m$ = 1.2\GeV. To study
 possible reflections from the $\phi \PKp$ spectrum, we consider $\phi \PKp$
 resonances with various masses, widths, and helicity angle distributions,
 but are not able to reproduce the pattern of structures seen in the
 $\JPsi\phi$ spectrum. Moreover, we separately analyze  the
 $\JPsi\phi$ spectrum for values of the $\phi \PKp$ masses larger than 1.9\GeV,
 a region of the Dalitz plot unaffected by postulated $\phi \PKp$ resonances,
 and  still observe  the
structure  near $\Delta m = 1.2$\GeV.

There are no candidate $\JPsi \PKp$ resonances reported in the literature.
 Still, we have considered such  resonances with various masses,
 widths, and helicity angle distributions.
No combination produces a reflected spectrum that matches
the observed $\JPsi \phi$ spectrum.

We have also checked the events with $\Delta m$ larger than 1.568\GeV that had been
eliminated from the analysis to ensure that they could not
cause similar reflections in the low-$\Delta m$ region.
After subtraction of the $\PBzs$ background the $\Delta m$
distribution of events with  $\Delta m$ larger than 1.568\GeV
is consistent with the prediction based on the three-body phase-space
hypothesis for the non-resonant background.
The extended $\Delta m$ plot is shown in Fig.~6 of \suppMaterial, and the corresponding 
fitted numbers of  $\PBp\to\JPsi\phi \PKp$ events for the 7 bins 
from the previously eliminated $\Delta m$ region are displayed in Fig.~7 of \suppMaterialSecond,  
 after subtracting the expected \PBzs background from simulation. 
Both distributions are consistent with the extrapolation from 
three-body phase-space.

\begin{figure}[!htb]
  \begin{center}
   \includegraphics[width=\linewidth]{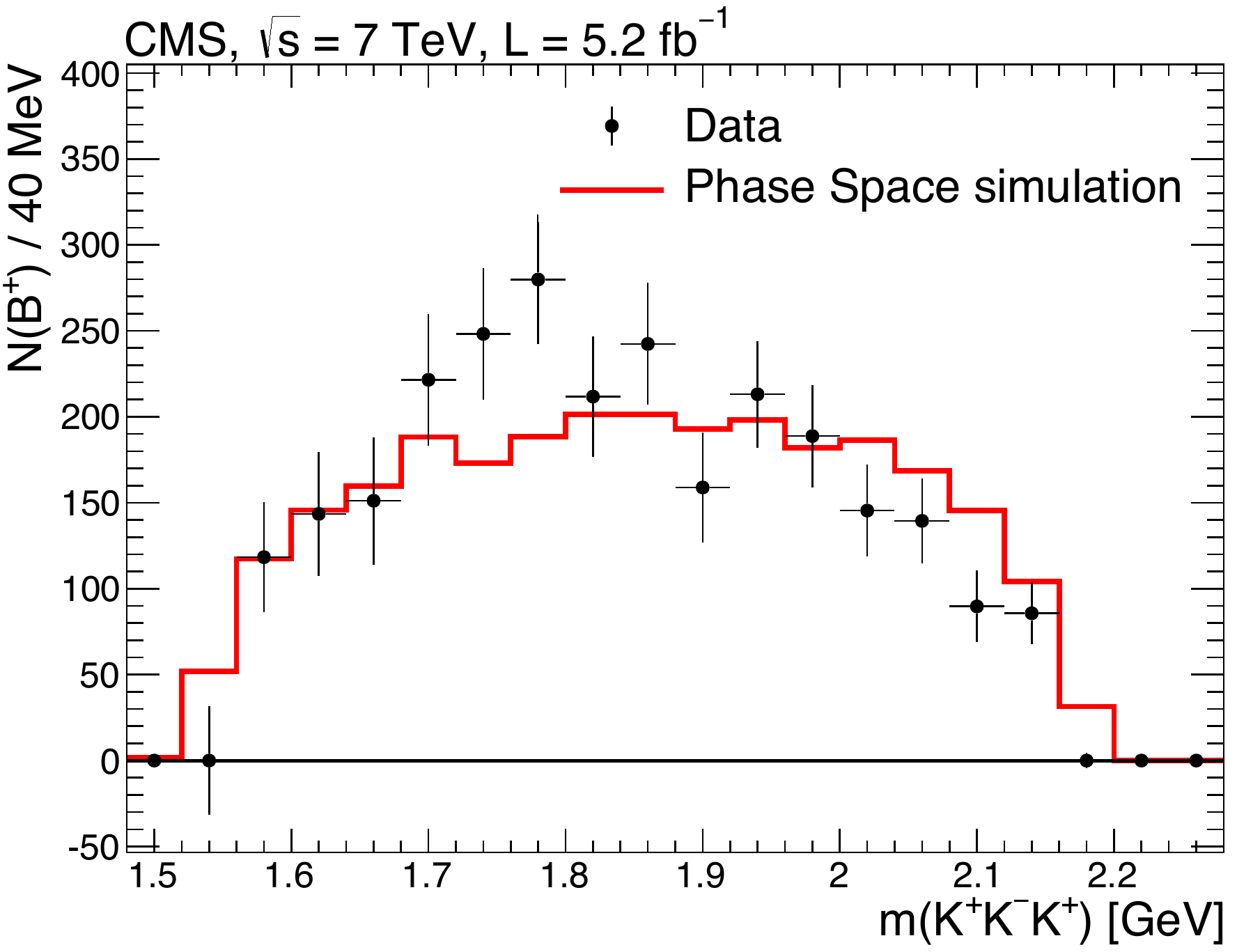}
    \caption{
The yield of $\PBp\to \JPsi\PKp \PKm \PKp$
candidates in the data as a function of the  $\PKp\PKm\PKp$
invariant mass. The error bars represent the statistical uncertainties.
The solid curve is the prediction from the PS simulation.
    }
    \label{fig:phikmass}
  \end{center}
\end{figure}

\begin{figure*}[!htb]
  \begin{center}
   \includegraphics[width=\linewidth]{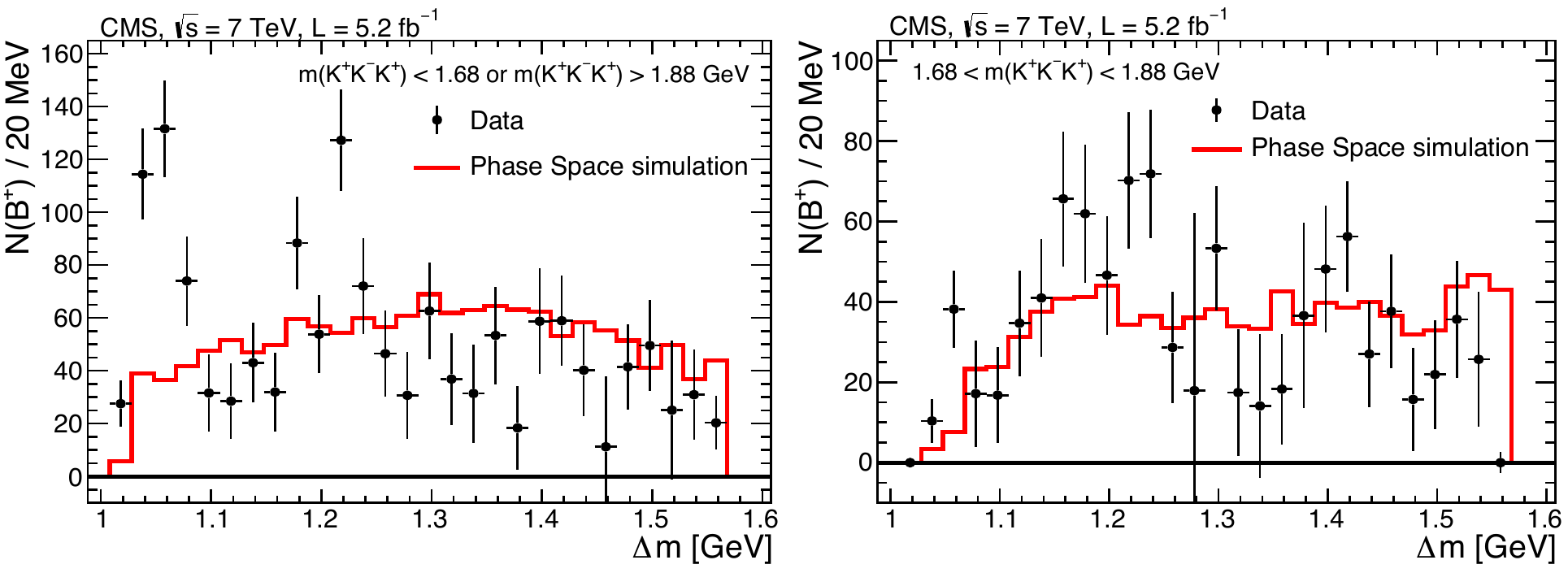}
    \caption{
The  number of  $\PBp\to \JPsi \phi \PKp$ candidates  as a
function of $\Delta m $    requiring either $m(\phi \PKp)<1.68$\GeV
or  $m(\phi \PKp)>1.88$\GeV (left), or $1.68<m(\phi \PKp)<1.88$\GeV (right).
The solid curve is the prediction from the PS simulation.
    }
    \label{fig:deltamless}
  \end{center}
\end{figure*}

 The results of these studies make it improbable that
 the two structures seen in
 the $\JPsi\phi$ spectrum are solely caused by reflections from
 resonances in the other two-body systems.  However, we cannot entirely exclude
 the possibility of such resonances.  For instance,
the $\PKp \PKm \PKp$ spectrum
 shown in Fig.~\ref{fig:phikmass} displays an excess of events
above the predicted
 PS distribution in the 1.7--1.8\GeV region, an excess that
 cannot be attributed to the presence of the $\JPsi\phi$ structure near
 threshold.   Figure~\ref{fig:phikmass} is obtained by dividing
the  $\JPsi\phi \PKp$ candidates  into 40\MeV-wide
$\PKp \PKm \PKp$ mass intervals
and fitting the  $\JPsi\phi \PKp$ invariant-mass distributions
for each interval  to extract
the $\PBp$ signal yield in that interval.
The $\Delta m $  distribution after excluding the
region ($1.68<m(\PKp \PKm \PKp)<1.88$\GeV) with the excess of events
is shown in the left plot of Fig.~\ref{fig:deltamless} and the corresponding distribution for
the excluded $\Delta m$ region in the right plot.
The presence of the lower-mass structure  is still apparent in the left plot, while
that of the higher-mass structure
is reduced though still visible.
Possible interference effects over the Dalitz plot could
 therefore distort the shape of the observed $\JPsi\phi$ structures
 and  affect the extraction of the resonance parameters.
The event sample is not
large enough to
 investigate these effects further.
 We assume that any interference effects can
 be neglected.
The structures in the $\JPsi \phi$ mass spectrum are described in terms of
zero, one, or two noninterfering resonances and   a nonresonant continuum
component.

We fit the two structures with $S$-wave relativistic BW  functions
convolved with a Gaussian mass resolution
function whose width varies linearly from 1\MeV at  threshold to about
4\MeV at $\Delta m$ = 1.25\GeV, as determined from simulation.
Each structure is described by a mass, width, and yield, all determined
from the fit.
The continuum is assumed to follow a three-body PS shape.
As an alternative, to check the sensitivity of the result to this assumption,
the shape of the continuum is obtained from an
event-mixing technique where the $\JPsi$, $\phi$, and $\PKp$ candidates are
selected from different events. We use two versions of the event mixing,
which differ by the $\phi$  and $\PKp$ candidates being selected in the same
event or not; they lead to almost identical shapes.
The differences between the two event-mixing shapes and
the three-body PS are used to evaluate
the systematic uncertainties in the continuum modeling.
To further investigate the effect of a possible $\phi\PKp$ resonance around 1.7\GeV as shown in Fig.~\ref{fig:phikmass},
we reweight our phase-space MC events with a $\phi\PKp$ mass distribution corresponding to a BW with a mass of 1.773\GeV
and a width of 200--300\MeV~\cite{Beringer:1900zz}. The helicity angle in the $\phi\PKp$ system is then weighted
to correspond to several different  assumptions about the decay of the possible resonance.
We estimate the yield of the possible $\phi\PKp$  resonance
in Fig.~\ref{fig:phikmass} to be  10\% of the total number of events.
We find that the shape of the PS $\Delta m$ distribution is always above the various distributions obtained from the above mixing
in the range  $\Delta m < 1.12$\GeV.
Thus, we conclude that using the PS distribution as the default background curve is more conservative with
respect to the significance of the low-mass peak if there is a possible effect from a $\phi\PKp$ resonance.

The masses and widths of the two structures are extracted by dividing the
 $\JPsi \phi \PKp$ candidates into 20\MeV-wide intervals
of  $\Delta m$ from 1.008 to 1.568\GeV and performing a global
unbinned maximum-likelihood (UML) fit to the $\JPsi \phi \PKp$
invariant-mass distribution in each $\Delta m$ interval.
The two data sets are fitted separately,
with a total of 56 mass spectra fitted simultaneously.
In each fit, the $\PBp$ mass is fixed to its nominal value
and the mass resolution $\delta$ is calculated using:
\begin{equation*}
 \delta = a_0 + a_1   \langle\Delta m\rangle  +a_2 \langle\Delta m\rangle ^2,
\end{equation*}
where $\langle\Delta m \rangle$ is the value of $\Delta m$  at the center of the bin, and
$a_0$, $a_1$, and $a_2$ are determined from simulation, separately for the two data sets.
The combinatorial background in each
bin is modeled as a second-degree  polynomial.
In the global fit,
the $\PBp$ yield is expressed as   the product
of  the relative efficiency times the number of signal events from the
two BWs and the
nonresonant continuum events. We fit  the $\JPsi \phi \PKp$
invariant-mass distribution for each $\Delta m$ bin from the two data sets
simultaneously by projecting the above product into each bin.
The  UML fit  returns signal  event yields of $310 \pm
70\stat$  and  $418 \pm 170\stat$ for
the lower- and higher-mass structures, respectively.
The corresponding mass difference and width values are:
$\Delta m_1=\massdiffone,
~\Gamma_1=\widthone; ~~\Delta m_2=\massdifftwo, ~\Gamma_2=\widthtwo$.
The projection of the UML fit assuming two structures  onto the $\JPsi \phi$ mass
spectrum is represented as the solid  line in Fig.~\ref{newfig3}.

As a check on the fitting procedure, we perform an
alternative one-dimensional (1D) binned $\chi^2$ fit to the
$\Delta m$  spectrum  shown in Fig.~\ref{newfig3}.
The same signal and background functions are used in the 1D fit as in
the global fit.
The result of the 1D fit, assuming two structures,
is shown  as the dashed  line in Fig.~\ref{newfig3}.
The measurements of the masses, widths, and yields of the two structures
from the global and 1D fits are in good agreement.

To evaluate the significance of each of the two  structures, three
UML  and  three 1D (binned $\chi^2$) fits are
performed on the data shown in Fig.~\ref{newfig3}:
(1)~a background-only fit (null-hypothesis);
(2)~a background plus a single $S$-wave relativistic BW signal function
convolved with a Gaussian resolution
function having a width   of 2\MeV for the lower-mass structure;
and (3)~a background plus two $S$-wave relativistic BW
functions  convolved with a Gaussian resolution function
to model both structures.
The log-likelihood ratio   $-2\Delta \ln{\mathcal{L}}$ in the case of the UML fits
or the $\chi^2$ change $\Delta\chi^2$ for the 1D fits between 1) and 2) is then a
measure of the statistical significance of the lower-mass structure, while the
corresponding values between fits (2) and (3) give a measure of the
statistical significance of the higher-mass structure.
The resulting values for a decrease in dof of 3
are  $-2\Delta \ln{\mathcal{L}}$ = 58
and  $\Delta\chi^2$ = 53 for the lower-mass structure,
and 36 and 37 for the higher-mass structure.

Simulated samples are
used to estimate the probability that background
fluctuations alone could give
rise to a signal as significant as that seen in the data for the lower-mass structure.
Over  50 million $\Delta m$ spectra were generated between 1.008 and 1.568\GeV with 2300 events for each spectrum
based on a three-body PS shape.
The most significant fluctuation in each spectrum is found
whose $\JPsi\phi$ invariant mass is
within $\pm 3$ times  the uncertainty in the CDF  mass value of 4.140\GeV
and having a  width between  10\MeV (half  the $\Delta m$ bin width)
to 80\MeV (half the separation between the two structures).
We then obtain the $\Delta \chi^2$ distributions in the simulated pure background samples and
compare  them with the corresponding value of the signal
in the data.
No generated spectrum  is found with a fluctuation having a $\Delta \chi^2$
greater than or equal to
the value obtained in the data ($53$).  The resulting $p$-value, taken as the fraction of the simulated samples
with a $\Delta \chi^2$ value greater than or equal to the value obtained in the data,
is  less than $2 \times 10^{-8}$, which corresponds to   a significance of more than 5
standard deviations.
Because the  second structure could be affected by  possible $\phi \PKp$ resonances,
it is difficult to model the background shape
in that mass range, and we do not quote  a numeric significance for the higher-mass
structure.
However, there is clear evidence
for a second structure around $\Delta m=1.2$\GeV even after excluding
the region with  possible $\PK_2$ resonances.
There is also a small excess of events around $\Delta m=1.4$\GeV, but with a local
significance of less than 3 standard deviations.

Various checks are made to examine the robustness of the two structures.
Each selection criterion is individually varied,  and in no case is
there an indication of a bias in the selection procedure.
The relative efficiencies for the first five $\Delta m$ bins are varied
by $\pm$20\% and the fit repeated,  confirming the robustness
of the significance of the first structure.
The $\Delta m$ distribution from an sPlot~\cite{Pivk:2004ty} projection
is compared to  the $\Delta m$ distribution shown in
Fig.~\ref{newfig3}.  No indication of bias is found.
The sPlot algorithm is a background-subtraction technique that
weights each event based on
the observed signal-to-background ratio, in this case from the fit to
the $\JPsi\phi \PKp$ mass distribution  shown in Fig.~\ref{fig:bmass}.
We repeat the analysis with the tighter requirements discussed earlier that
lower the combinatorial
background level by a factor of ten and retain 40\% of the $\PBp$ events,
as shown in the right plot in Fig.~\ref{fig:bmass}.
The $\Delta m $ plot for these events looks similar to Fig.~\ref{newfig3},
showing two peaking structures whose fitted mass and width values are consistent with the
results from the nominal data sample.
No indication of a possible bias is found.

The estimations of the contributions to the  systematic uncertainties in
the mass and width measurements of the two structures
shown in Table~\ref{table:SysVariations}, are determined from several studies.
 The uncertainties owing to the probability density functions (PDFs) for
the  combinatorial background shape in the $m(\JPsi\phi \PKp)$ spectrum
and the $\PBp$ signal
are studied by using different PDFs  such as first- and  third-degree polynomials,
exponential functions, and a number of Gaussian functions.
The uncertainties in the shape of the
relative efficiency vs.\ $\Delta m$ are evaluated by varying  the relative efficiency in various bins
and comparing with the 2D efficiencies for correction of $m(\JPsi\phi)$ vs.\ $m(\phi \PKp)$.
 The uncertainties caused by the binning of the $\Delta m$ spectrum are studied by
using  10\MeV bins instead of 20\MeV bins.
To estimate the uncertainty from the signal fitting function,
 we repeat the fit to
the $\Delta m$ distribution using either a nonrelativistic BW  or a $P$-wave relativistic BW function for
each structure.
 The uncertainties from the $\Delta m$ mass resolution are studied by varying the  mass resolution values
obtained from simulation within their statistical uncertainties.
 To evaluate potential distortions in the $\Delta m$ background shape
caused by possible  $\phi \PKp$ resonances, we obtain the $\Delta m$ background
shape from data using an event-mixing technique
by applying the same kinematic constraints and taking the $\phi$ and $\PKp$
candidates from the same event,
but the  $\JPsi$ candidate from a different event.
The uncertainties due to selection requirements are studied
in the MC sample.
The overall systematic uncertainties in the measurement of the masses and
widths of the two structures are found by adding in quadrature the individual
combinations summarized in Table~\ref{table:SysVariations}.

\begin{table*}[htbp]\topcaption{
Systematic uncertainties in the measured
masses and widths of the two peaking structures from the sources listed
and the total uncertainties.
}
\label{table:SysVariations}
	\begin{center}
		\begin{tabular}{l|cccc}
			\hline
	         &  ${\mathrm{m_1}}$  (\MeVns{}) & ${\Gamma_1}$ (\MeVns{}) &  ${\mathrm{m_2}}$  (\MeVns{}) & ${\Gamma_2}$  (\MeVns{})\\
			\hline
		  	$\PBp$ background PDF  & 0.8  & 7.4  & 2.6  & 9.9  \\
                        $\PBp$ signal PDF  & 0.2  & 3.6  & 2.7  & 0.2  \\
                        Relative efficiency & 4.8 & 6.0 & 0.9 & 10.0 \\
			$\Delta m$ binning & 3.7  & 1.5  & 2.7  & 0.2  \\
			$\Delta m$ structure  PDF    & 0.8  & 9.3  & 0.6  & 4.9  \\
			$\Delta m$ mass resolution & 0.8  & 6.4  & 0.6  & 4.6  \\
		        $\Delta m$ background shape    & 0.2  & 7.0  & 0.3  & 0.2  \\
                        Selection requirements  & 0.8  & 7.8  & 5.5  & 1.8  \\
			\hline
		        Total          & 6.3  & 19  & 7.3  & 16  \\
			\hline
		\end{tabular}
	\end{center}
\end{table*}

\section{Summary}

In summary, a peaking structure in the $\JPsi\phi$  mass spectrum from
$\PBp\to \JPsi \phi \PKp$ decays has been observed
in pp collisions at $\sqrt{s}$ = 7\TeV by the CMS Collaboration at
the LHC.
Assuming an $S$-wave relativistic BW
lineshape for this structure above a three-body PS shape for the nonresonant
background, a statistical significance of greater than
5 standard deviations is found.
Adding the $\JPsi$ mass~\cite{Beringer:1900zz} to the extracted $\Delta m$ values,
the mass and width    are measured to be
$m_1=\massonewithsys$ and $\Gamma_1=\widthonewithsys$.
The measured mass and width are consistent with the $\mathrm{Y}(4140)$ values reported by CDF experiment.
The relative branching fraction of this peaking structure with respect to  the total number
of $\PBp\to \JPsi \phi \PKp$ events is estimated
to be about 0.10, with a statistical uncertainty of about 30\%. This is consistent with both the value measured by CDF of $15\% \pm 5\%$
and  the upper limit  reported by LHCb (0.07).
In addition,  evidence for  a second peaking structure is found in the same mass
spectrum, with measured  mass and width values  of
$m_2=\masstwowithsys$ and    $\Gamma_2=\widthtwowithsys$.
Because of possible reflections from  two-body decays,
the statistical significance of the second structure cannot be
reliably determined.
The two structures are well above the threshold of open charm
($\PD\PaD$)
decays and have relatively narrow widths.
Conventional charmonium mesons with these masses would be expected
to have larger widths and to decay predominantly into open charm pairs  with
small branching fractions into $\JPsi\phi$.
Angular analyses of the $\PBp\to \JPsi \phi \PKp$
decays would help  elucidate the nature of these structures.

We congratulate our colleagues in the CERN accelerator departments for the excellent performance of the LHC and thank the technical and administrative staffs at CERN and at other CMS institutes for their contributions to the success of the CMS effort. In addition, we gratefully acknowledge the computing centres and personnel of the Worldwide LHC Computing Grid for delivering so effectively the computing infrastructure essential to our analyses. Finally, we acknowledge the enduring support for the construction and operation of the LHC and the CMS detector provided by the following funding agencies: BMWF and FWF (Austria); FNRS and FWO (Belgium); CNPq, CAPES, FAPERJ, and FAPESP (Brazil); MES (Bulgaria); CERN; CAS, MoST, and NSFC (China); COLCIENCIAS (Colombia); MSES (Croatia); RPF (Cyprus); MoER, SF0690030s09 and ERDF (Estonia); Academy of Finland, MEC, and HIP (Finland); CEA and CNRS/IN2P3 (France); BMBF, DFG, and HGF (Germany); GSRT (Greece); OTKA and NKTH (Hungary); DAE and DST (India); IPM (Iran); SFI (Ireland); INFN (Italy); NRF and WCU (Republic of Korea); LAS (Lithuania); CINVESTAV, CONACYT, SEP, and UASLP-FAI (Mexico); MBIE (New Zealand); PAEC (Pakistan); MSHE and NSC (Poland); FCT (Portugal); JINR (Dubna); MON, RosAtom, RAS and RFBR (Russia); MESTD (Serbia); SEIDI and CPAN (Spain); Swiss Funding Agencies (Switzerland); NSC (Taipei); ThEPCenter, IPST, STAR and NSTDA (Thailand); TUBITAK and TAEK (Turkey); NASU (Ukraine); STFC (United Kingdom); DOE and NSF (USA).

Individuals have received support from the Marie-Curie programme and the European Research Council and EPLANET (European Union); the Leventis Foundation; the A. P. Sloan Foundation; the Alexander von Humboldt Foundation; the Belgian Federal Science Policy Office; the Fonds pour la Formation \`a la Recherche dans l'Industrie et dans l'Agriculture (FRIA-Belgium); the Agentschap voor Innovatie door Wetenschap en Technologie (IWT-Belgium); the Ministry of Education, Youth and Sports (MEYS) of Czech Republic; the Council of Science and Industrial Research, India; the Compagnia di San Paolo (Torino); the HOMING PLUS programme of Foundation for Polish Science, cofinanced by EU, Regional Development Fund; and the Thalis and Aristeia programmes cofinanced by EU-ESF and the Greek NSRF.

\bibliography{auto_generated}   
    \ifthenelse{\boolean{cms@external}}{}{
    \clearpage
    \appendix
    \section{Supplemental Material\label{app:suppMat}}
    
Figure \ref{fig:rawdm} shows the continuation of the  $\Delta m$ spectrum for $\Delta m >1.568$\GeV, including the
contribution from non-B candidates, after subtracting the expected  \PBzs   contribution from
simulation for candidate events with $\JPsi\phi \PKp$ invariant mass
within ${\pm}1.5\sigma$ ($\sigma=9.3$\MeV) of the B nominal mass.  Figure~\ref{fig:wholedm}
shows the extension of the $\Delta m$ spectrum in Fig.~3 in the paper,
excluding non-B background, to the full phase space.
The absence of strong activity in the high-$\Delta m$ region reinforces our conclusion that
the near-threshold narrow structure is not due to a reflection of other resonances.

\begin{figure*}[htb]
\centering
   \includegraphics[width=\textwidth]{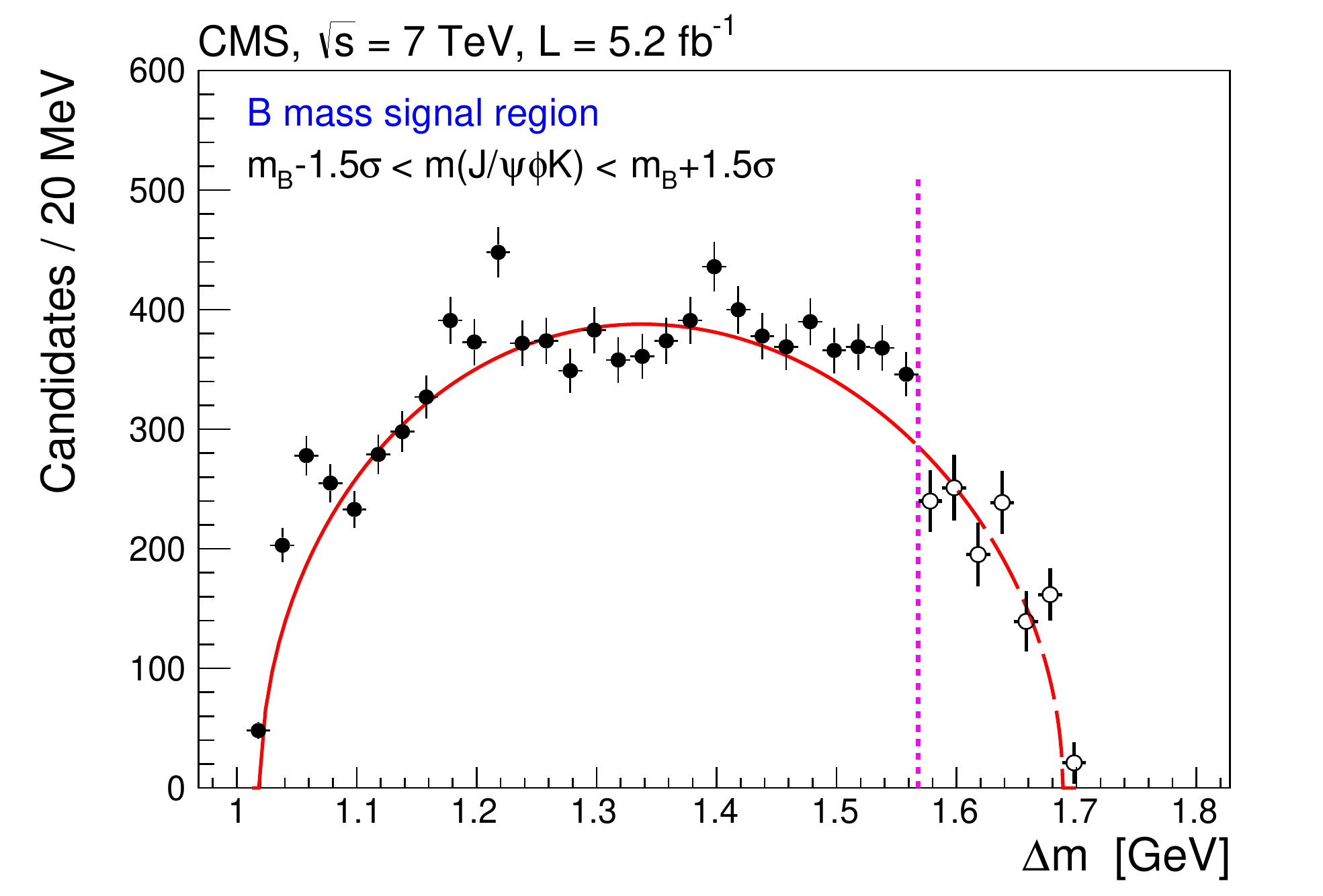}
    \caption{
The $\Delta m$ spectrum, including non-B candidates after subtracting the expected \PBzs  contribution from
simulation for candidate events with $\JPsi\phi \PKp$ invariant mass within ${\pm}1.5\sigma$ ($\sigma=9.3$\MeV) of
the B nominal mass. The dashed vertical line indicates the boundary of the region eliminated from the
analysis (the region to the right of the dashed line).
The data points in the region to the left of the dashed line  are from events used in the analysis and are represented
by filled circles  with error bars. The open circles  in
the region to the right of the dashed line are the result of  subtracting  the expected \PBzs  background
from simulation, and their uncertainties are correlated.
The solid curve is the prediction for a three-body phase-space distribution, normalized to the total number of events
in the left region, after subtracting the
yields from the two low-mass peaking structures. The extrapolation of the phase-space prediction into the right-hand
region is shown by the dashed curve.
}
    \label{fig:rawdm}
\end{figure*}

\begin{figure*}[htbp]
\centering
   \includegraphics[width=\textwidth]{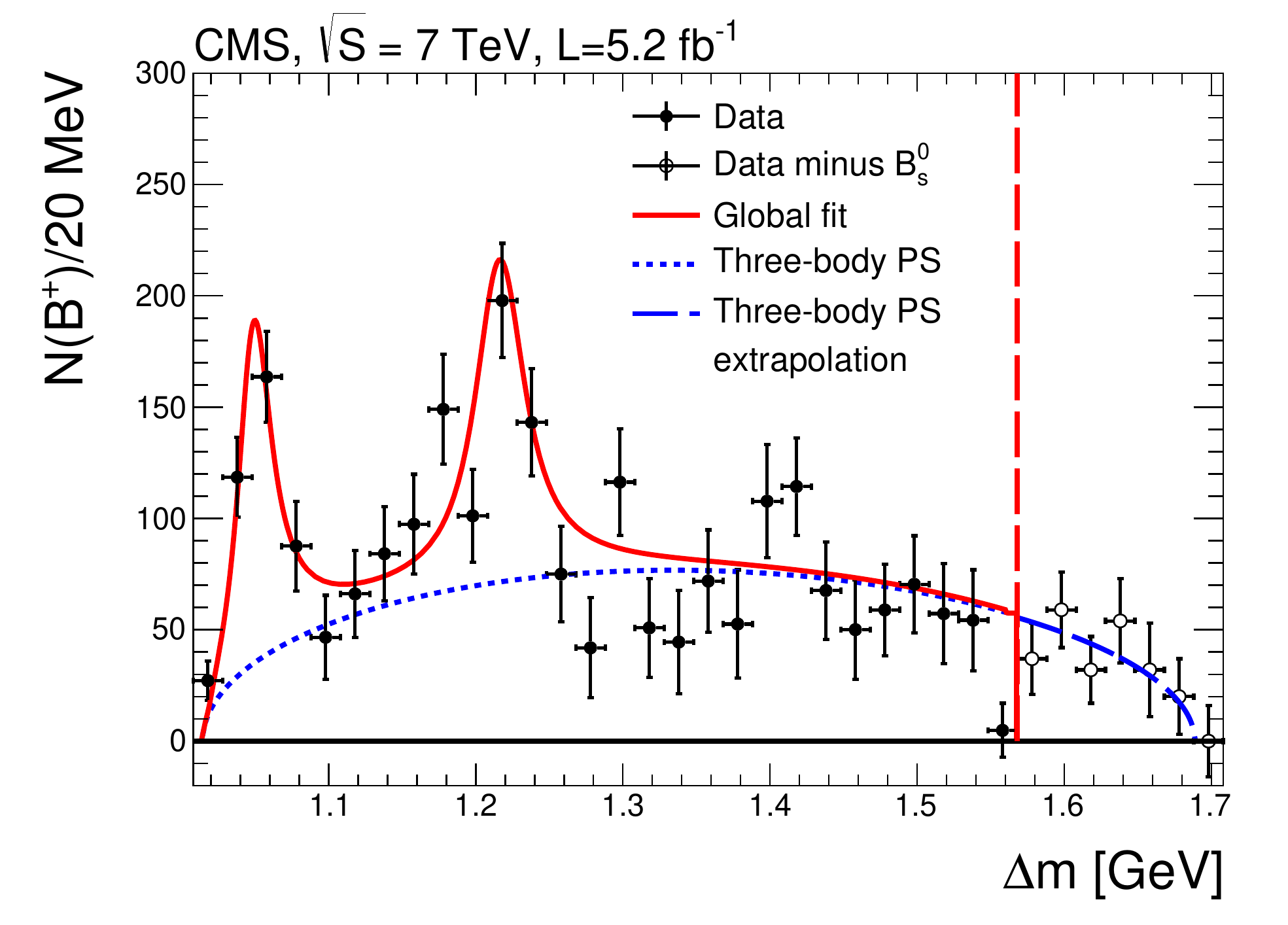}
    \caption{
The number of $\mathrm{B}^+\rightarrow \JPsi \phi \mathrm{K}^+$ candidates as a function of $\Delta m$.
The filled circles to the left of the vertical dashed line are results used in the analysis.
The solid curve is the result of the global fit described in the text. The points and the curve
are a repeat of those in Fig.~3. The open circles to the right of the vertical dashed line are extracted in the
seven bins of the previously eliminated $\Delta m$ region, after subtraction of the predicted
 \PBzs background from simulation.
The short-dashed curve represents the prediction from three-body phase-space for the nonresonant background,
normalized to the number of expected background events in the region to the left of the vertical dashed line.
The long-dashed curve is an  extrapolation of this prediction into the previously eliminated region.
}
    \label{fig:wholedm}
\end{figure*}

    }
\cleardoublepage \section{The CMS Collaboration \label{app:collab}}\begin{sloppypar}\hyphenpenalty=5000\widowpenalty=500\clubpenalty=5000\textbf{Yerevan Physics Institute,  Yerevan,  Armenia}\\*[0pt]
S.~Chatrchyan, V.~Khachatryan, A.M.~Sirunyan, A.~Tumasyan
\vskip\cmsinstskip
\textbf{Institut f\"{u}r Hochenergiephysik der OeAW,  Wien,  Austria}\\*[0pt]
W.~Adam, T.~Bergauer, M.~Dragicevic, J.~Er\"{o}, C.~Fabjan\cmsAuthorMark{1}, M.~Friedl, R.~Fr\"{u}hwirth\cmsAuthorMark{1}, V.M.~Ghete, N.~H\"{o}rmann, J.~Hrubec, M.~Jeitler\cmsAuthorMark{1}, W.~Kiesenhofer, V.~Kn\"{u}nz, M.~Krammer\cmsAuthorMark{1}, I.~Kr\"{a}tschmer, D.~Liko, I.~Mikulec, D.~Rabady\cmsAuthorMark{2}, B.~Rahbaran, C.~Rohringer, H.~Rohringer, R.~Sch\"{o}fbeck, J.~Strauss, A.~Taurok, W.~Treberer-Treberspurg, W.~Waltenberger, C.-E.~Wulz\cmsAuthorMark{1}
% ...
\vskip\cmsinstskip
% ...
1:~~Also at Vienna University of Technology, Vienna, Austria\\
2:~~Also at CERN, European Organization for Nuclear Research, Geneva, Switzerland\\
% ...

\end{sloppypar}
\end{document}